\documentclass[doublecol]{epl2}

\title{Pore opening effects and transport diffusion in the Knudsen regime in comparison to self- (or tracer-) diffusion}
\shorttitle{Pore opening effects and transport diffusion in the Knudsen regime} 

\author{S. Zschiegner\inst{1,2} \and S. Russ\inst{1} \and A. Bunde\inst{1} \and J. K\"arger\inst{2}}
\shortauthor{S. Zschiegner, S. Russ, A. Bunde, and J. K\"arger}  

\institute{       
	\inst{1} Universit\"at Giessen, Institut f\"ur Theoretische Physik III, Germany \\
	\inst{2} Universit\"at Leipzig, Fakult\"at f\" ur Physik und Geowissenschaften, Germany
}

\abstract{We study molecular diffusion in linear nanopores with different types of roughness in the so-called Knudsen regime. Knudsen diffusion represents the limiting case of molecular diffusion in pores, where mutual encounters of the molecules within the free pore space may be neglected and the time of flight between subsequent collisions with the pore walls significantly exceeds the interaction time between the pore wall and the molecules. 
We present an extension of a commonly used procedure to calculate transport diffusion coefficients. Our results show that using this extension, the coefficients of self- and transport diffusion in the Knudsen regime are equal for all regarded systems, which improves previous literature data.}

\begin{document}

\maketitle 

\section{Introduction}
Diffusion phenomena of gases in disordered and porous 
media have been subject to intense research for several decades 
\cite{keil,kaerger92,chen,coppens1} with applications in heterogeneous 
catalysis \cite{ertl}, adsorption \cite{schueth} and separation \cite{krishna}. 
Recent progress in synthesizing nanostructured porous materials 
\cite{schueth,kaevalvas} has provided essentially unlimited options for the 
generation of purpose-taylored pore architectures and there is an increasing demand for clarification of the main features of molecular transport in such systems \cite{anandan,caro}. 
In this work, we concentrate on transport pores, that play an important role in bimodal porous materials \cite{bimodal} where they ensure fast molecular exchange between the microporous regions, in which the actual conversion and separation phenomena take place. In the transport pores, the so-called Knudsen diffusion dominates, where the intermolecular collisions can be neglected and the molecules perform a series of free flights and change direction statistically after collisions with the pore walls.  

\begin{figure}
\onefigure[width=0.47\textwidth]{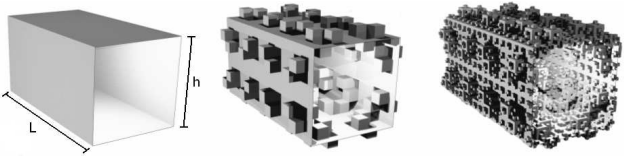} 
\caption{Realizations of the pore geometry generated by a generalized random $3d$-Koch curve of generation $\nu$. (a) Smooth pore ($\nu=0$) and (b-c) rough pores of $\nu=1$ and $2$ of length $L$ and width $h$ (for details see \cite{russprerap}). For the simulations, the system is covered with a grid of lattice constant $a=h/64$.}
\label{f.3Dpores}
\end{figure}

Experimentally, two kinds of diffusion problems can be considered, the 
so-called transport diffusion, where the particles diffuse in a 
non-equilibrium situation from one side of the system to the opposite side 
(here under the influence of a concentration gradient) and the self- (or 
tracer-) diffusion under equilibrium conditions. Both problems are 
described by the transport diffusion coefficient $D_t$ and the self- 
(or tracer-) diffusion coefficient $D_s$, respectively. $D_t$ is defined by Fick's 1st law as
the proportionality constant between the current density $j$ and the concentration gradient $\partial c/\partial{x}$,
\begin{equation}
\label{e.j}
j=-D_{t}\frac{\partial{c}}{\partial{x}},
\end{equation}
while $D_s$ is defined by the mean square displacement ${<x^{2}(t)>}$ of a random walker
\begin{equation}
\label{e.xsquares}
<x^{2}(t)>=2dD_{s}t
\end{equation}
after time $t$, where $d$ is the dimension of the pore.
Under the conditions of Knudsen diffusion, both diffusion coefficients are expected to coincide \cite{kaerger92}. Using the method of Evans \textit{et al} \cite{evans}, this has recently been verified numerically for smooth pores and for pores of different surface roughness \cite{russprerap} showing, however, slight numerical differences between $D_s$ and $D_t$. 
Since, on the other side, there are also studies which do in fact consider the possibility of a difference between transport and self-diffusion coefficients in the Knudsen regime \cite{coppens2}, even these differences have to be taken seriously. In \cite{russprerap} their occurence is attributed to the possibility of a non-constant density gradient of the gas particles inside the pores that disturbs the method of \cite{evans} slightly. Here, we investigate this problem in detail both for diffusion in the pores and for quite general random walks. We find that the deviations from the expected gradient are strongest when the particles perform discrete jumps with many different velocities. The deviations disappear completely, if only one fixed jump length per time step is allowed and they are quite small (but measurable) for the Knudsen pores of Ref.~\cite{russprerap} (see Fig.~\ref{f.3Dpores}). We present an improvement of Evan's method (that we call "Enhanced $f_t$ Method" (EFM)) that accounts for the non-constant gradients and show that with EFM, $D_s=D_t$ for all considered systems. 

\section{Calculation of the transport diffusion}

In the simulations of $D_t$ as well as in a typical diffusion experiment, a concentration gradient $\partial c(x)/\partial x$ is applied with the concentrations $c=c_{0}=1$ on the left hand side and $c=0$ on the right hand side ($x\geq L$) of the pore, respectively. All particles start at $x=0$ and perform a random walk in $d=1$ or a random trajectory between the system walls in $d=3$, using Lambert's law of reflection \cite{russprerap}. They are absorbed when they hit the left or the right boundary (Dirichlet boundary conditions). 
After some relaxation time, this leads to a constant current density $j$, described by Eq.~(\ref{e.j}). 

Relaxation of a particle flow into a stationary state is very time-consuming, as the particle flux has to be monitored throughout the system. 
In the stationary state the particle concentration does not depend on the relaxation time $t$. In our simulations we consider a state as stationary if the particle concentrations between two given time steps differ at maximum by a predetermined threshold value \cite{footnote1}. 
It is common practice to use the faster Evans' method \cite{evans} to derive $j$ from the (transmission) probability $f_{t}$ that a particle starting at the left boundary will leave the pore through the right boundary,
\begin{equation}
\label{e.jft}
j=c_{0}f_{t}<u_{x}>
\end{equation}
where $<u_{x}>$ is the mean velocity in $x$-direction. Combining Eqs.~(\ref{e.j}) and (\ref{e.jft}) yields
\begin{equation}
\label{e.Dtrichtig}
D_t=-c_{0}f_{t}<u_{x}>\left(\frac{dc}{dx}\right)^{-1}.
\end{equation}
Usually, the concentration gradient is assumed to be constant and equal to
\begin{equation}
\label{e.dcdx}
\frac{\partial{c}}{\partial{x}}=-c_{0}/L.
\end{equation}	 
Combining Eqs.~(\ref{e.Dtrichtig}) and (\ref{e.dcdx}) yields
\begin{equation}
\label{e.dtft}
D_{t}= f_{t} <u_{x}> L.
\end{equation}
Accordingly, for calculating $f_{t}$, $N$ random trajectories are considered that start at $x=0$ and end when either $x=0$ or $x=L$ is reached.  
As we show in this paper, the problem with Evans' method is that it only works if the concentration gradient is well described by Eq.~(\ref{e.dcdx}). It leads to spurious results, if deviations from a constant concentration gradient occur and in this case, Eq.~(\ref{e.Dtrichtig}) has to be taken as starting point.

\begin{figure}
\onefigure[width=0.33\textwidth]{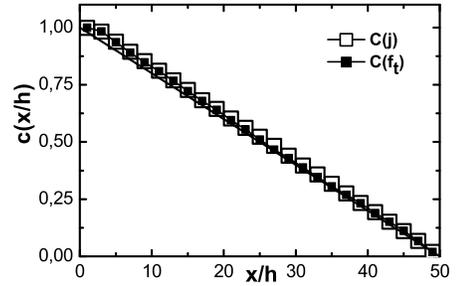}
\caption{Concentration densities $c(x/h)$ as calculated through relaxation of $10^6$ particles into a stationary state (open symbols) and by the Enhanced $f_t$ Method (EFM, filled symbols), plotted versus $x/h$ for smooth 3d pores of length $L=50h$. The results were generated from $10^6$ runs.
}
\label{f.3drelaxation}
\end{figure}

\section{From probability density to concentration}

To obtain the correct value of $D_t$ via Eq.~(\ref{e.Dtrichtig}), we need the concentration $c(x)$ within the pore and the associated concentration gradient $\partial c(x)/\partial x$. 
Figure~\ref{f.3drelaxation} shows the results for $c(x)$ as calculated by relaxing the density of gas particles inside smooth 3d pores of length $L=50h$ and width $h$ into a stationary state (open symbols). 

For the simulation $h$ was subdivided into smaller segments of size $a$
with $h=64a$. Therefore, $a$ is the lattice constant which for the
roughest pores considered is the size of the smallest boundary structures.
When the particle hits a wall, the new direction is chosen randomly,
according to Lambert's law of reflection. The path of the particle and hence
the next collision is calculated by using basic mathematics within successive
volume elements of size $a^3$. The traveled
distances are evaluated after all integer numbers of time steps $\tau$,
whereas for the path lengths, non-integer multiples of $a$ are allowed and
determined by linear interpolation.

We can see that the concentration profile in Fig.~\ref{f.3drelaxation} differs only inside a small boundary region from a linear profile. Close to the left boundary, $c(x)$ shows a small bump, whose relative position $x_{\rm{min}}/L$ approaches zero for increasing system size $L$. Therefore, we can still apply Evans' method for $x\gg x_{\rm{min}}$, where $\partial c(x)/\partial x=\rm{const}$ but unequal to $-1/L$. The deviation of the slope from $-1/L$ changes $D_t$ by several percent. 
Unfortunately, the relative deviation between the correct $\partial c(x)/\partial x$ and $-1/L$ is unaffected by the system size, as the absolute deviation decreases with the system size in the same way as $\partial c(x)/\partial x$. Hence, the problem cannot be solved by simply increasing the system size, but only by calculating the correct $c(x)$.

\begin{figure}
\onefigure[width=0.48\textwidth]{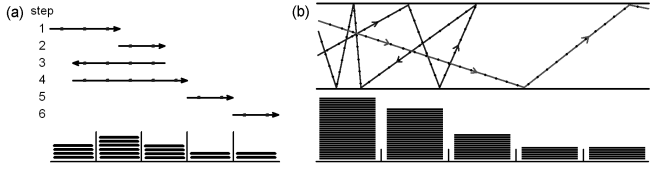}
\caption{Schematics for generating the particle histogram and hence $c(x)$ for (a) 1d jumps with constant velocity and (b) particle trajectories in 2D. Every time step, the histograms (below) are incremented at the appropriate location.  
}
\label{f.schematic}
\end{figure}

Therefore, if we want to apply Evans method, we must find a fast procedure to calculate the correct $c(x)$ for large systems. 
Figure~\ref{f.schematic} shows the schematics of such a procedure in $1d$ and $2d$, where $c(x)$ is calculated from the trajectories used to calculate $f_t$. 
The particle positions at every time step are indicated by dots. 
For better statistics we used $64$ time steps in every time interval $\tau$ that a particle with $u=1$ needs to travel the distance of $h$. This has to be taken into account when normalizing the concentration $c(x)$.
The histogram over all particle positions of the whole simulation is shown at the bottom of the figure. 
Hence, the histogram describes the time-averaged concentration $\widetilde{c}(x)$. Using the ergodic hypothesis we identify $\widetilde{c}(x)$ as equal to the ensemble averaged concentration $c(x)$ that we can therefore calculate directly from the particle histogram. 
To test this assumption, $c(x)$ of our three-dimensional pores calculated by EFM is shown in Fig.~\ref{f.3drelaxation} by the filled symbols. It can be seen that indeed $c(x)$ agrees very well with the profile calculated by relaxing the system into an equilibrium situation. The time-consuming relaxation into a stationary state therefore can still be avoided and the fast method of Ref.~\cite{evans} is maintained and improved.

\section{Examples and results}

\begin{figure}
\onefigure[width=0.48\textwidth]{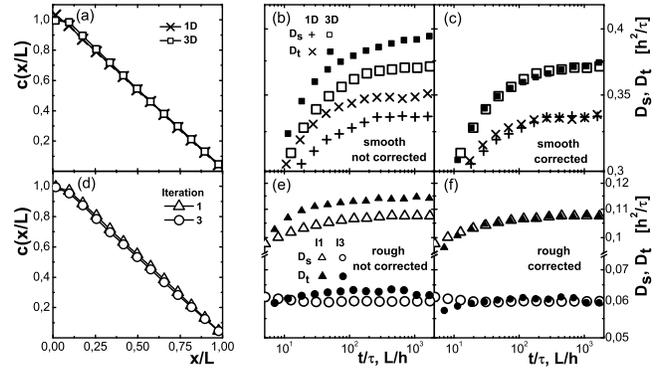}
\caption{
(a) Particle concentration $c(x/L)$ plotted vs. the normalized position $x/L$ along the $x$-axis for smooth 3d pores (squares) and the corresponding case of Levy distributed jumps in 1d (x). (b) Self- (open symbols, +) and transport (filled symbols, x) diffusion coefficients for the 3d pores and the 1d jumps without correction and (c) with correction. (d) Concentration profile and (e) uncorrected and (f) corrected self- and transport diffusion coefficients (open and filled symbols, respectively) for rough 3d pores of generation $\nu=1$ (triangles) and $\nu=3$ (circles). The results were generated from $10^7$ runs. 
}
\label{f.3Droughprofile}
\end{figure}

\begin{figure}
\onefigure[width=0.43\textwidth]{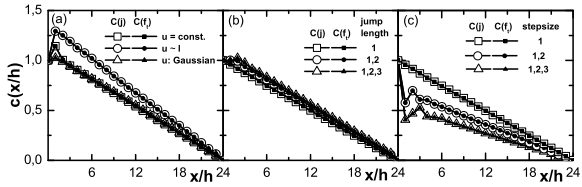}
\caption{Concentration densities $c(x/h)$ of several linear systems calculated by relaxation into a stationary state (open symbols) and by EFM (filled symbols), plotted versus $x$ for systems of length $L=24h$. (a) Gaussian distributed step sizes with different velocity profiles: constant velocity (squares), velocity proportional to the jump length (circles) and Gaussian distributed velocities (triangles). 
(b) Discrete random walks with fixed velocity $u=h/\tau$ and stepsizes $l=1h$ (squares), $1h,2h$ (circles) and $1h,2h,3h$ (triangles), (c) Same random walks but with velocity $u=l/\tau$. The expected  $c(x)=c_0-x/L$ only applies to the cases of $l=1h$ (in (b) and (c)).  The results were generated from $10^6$ runs.
}
\label{f.Gaussprofile}
\end{figure}

\begin{figure}
\onefigure[width=0.49\textwidth]{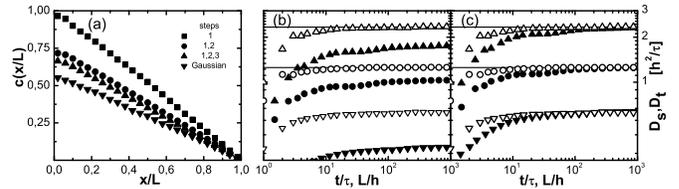}
\caption{(a) Concentration density $c(x/L)$ for random walks with discrete jump lengths $l=1h$ (squares), $l\in\{1h,2h\}$ (circles) and $l\in\{1h,2h,3h\}$ (triangles up) and Gaussian distributed $l$ (triangles down) with $\mu=0$ and $\sigma=2.5h$, plotted versus the normalized position $x/L$. Self- (empty symbols) and transport (filled symbols) diffusion coefficients for some of the 1d random walks from (a) without correction (b) and with correction (c). The straight lines show the values of $D_s$ and $D_t$ obtained analytically. The results were generated from $10^6$ runs.}
\label{f.RWprofile}
\end{figure}

In the following, we show $c(x)$ and the uncorrected and corrected diffusion coefficients $D_s$ and $D_t$ for several different systems, including smooth and rough linear 3d pores (Fig.~\ref{f.3Droughprofile}) and  different 1d random walks 
(Figs.~\ref{f.Gaussprofile} and~\ref{f.RWprofile}). All diffusion coefficients are in units of $h^{2}/\tau$ since we measure the length in units of $h$ and the time in units of $\tau$.
We show that using our new method, in all considered cases, both diffusion coefficients are in excellent agreement.

We start with the 3d pores as the experimentally relevant case (see Fig.~\ref{f.3Dpores}). We know from \cite{russprerap} that their jump lengths $l$ asymptotically show a Levy-distribution $P(\left|l/h\right|)\sim\left|l/h\right|^{-(1+\beta)}$ with $\beta=3$. Therefore we include simulations in 1d with a similar Levy jump length distribution with $\beta=3$ into the same figure. To make the simulations in 3d and 1d more comparable we used a composed jump length distribution in our 1d simulations: Jumps larger than $h$ are Levy distributed, jumps smaller than $h$ occur equally often. This is to imitate the influence of the pore geometry for small jump lengths. 
Figure~\ref{f.3Droughprofile} shows (a) the resulting concentration profiles $c(x)$ for these 1d and the smooth 3d systems, and 
the resulting diffusion coefficients (b) without correction and (c) with correction. Figures~\ref{f.3Droughprofile}(d)-(f) show the equivalent results for rough 3d pores of first and third generation. 
All deviations between $D_s$ and the uncorrected $D_t$ are about $5\%$ and we can see they disappear by using EFM. 

It is interesting that the deviations are smaller for the rough pores than for $\nu=0$. We believe that this is due to the very large jump lengths occuring in the smooth pores that are stronger suppressed by the boundary roughness. Since higher generations only influence smaller length scales this effect saturates for higher roughness. 
Additionally for small lengths $L$ and for small times $t$ we see huge finite size effects, but both $D_t$ and $D_s$ are converging asymptotical. In this regime the rough geometries show more statistical fluctuations since fewer particles diffuse as far into the pore.

Next, we show the calculations of $c(x)$ by EFM for some more theoretical cases of 1d random walks with different jump length and velocity distributions. 
In Fig.~\ref{f.Gaussprofile} we compare $c(x)$ for these cases as calculated by relaxation into a steady state (open symbols) to $c(x)$ as calculated by EFM (filled symbols). Also here, the agreement between both methods is excellent. 
In Fig.~\ref{f.Gaussprofile}(a) the jump lengths $l$ are Gaussian distributed, whereas the velocities $u$ are (i) constant, (ii) proportional to $l$ and (iii) Gaussian distributed.
Figure~\ref{f.Gaussprofile}(b) shows $c(x)$ for random walks of step size  $l=nh$, $n\in\{1,2,3\}$ with constant velocity $u=h/\tau$, where $u$ can be identified with the mean velocity in real systems.
Figure~\ref{f.Gaussprofile}(c) shows $c(x)$ for the same random walks but in this case per time step $\tau$ one complete jump is made and jumps of different length therefore have different velocities.

In all considered 1d systems, 
only for systems with a single jump length $l=\pm h$, $c(x)$ does not deviate from the expected linear concentration profile with $\partial c(x)/\partial x=-1/L$. 
A very pronounced deviation can be seen for systems with high probabilities to make either short or long jumps per time step. 
In this case, the concentration profile can be much higher or much smaller than in the other cases (Fig.~\ref{f.Gaussprofile}(a) and (c)). For constant velocities or for a Gaussian velocity profile which is not correlated to the jump lengths, the enhancement of the concentration profile is comparable to the results of the 3d pores.  

As a last example, Fig.~\ref{f.RWprofile}(a) shows the concentration profile for random walks with discrete step sizes of $l=1h$, $l\in\{1h,2h\}$, $l\in\{1h,2h,3h\}$, each with equal probability, and for Gaussian distibuted step sizes with mean $\mu=0$ and standard deviation $\sigma=2.5h$, all with constant $u$. 
Figure~\ref{f.RWprofile}(b) shows 
$D_s$ and $D_t$, where $D_t$ is calculated using Eq.~(\ref{e.dtft}). 
We see clearly that all cases of non-linear $c(x)$ lead to large deviations between both coefficients. Accordingly, we now correct $D_t$ by using $\partial c(x)/\partial x$ from EFM instead of Eq.~(\ref{e.dcdx}). The corrected results are shown in Fig.~\ref{f.RWprofile}(c). For the discrete cases, we also calculated $D_s$ and $D_t$ analytically, yielding $D_s=D_t$. These values are displayed in Fig.~\ref{f.RWprofile}(b,c) by straight lines and it can be seen that only $D_t$ as calculated by the uncorrected procedure deviates from the analytical value.

\begin{figure}
\onefigure[width=0.48\textwidth]{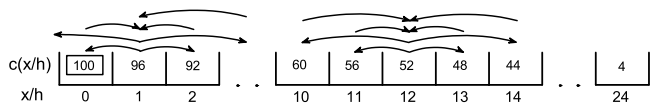}
\caption{Sketch of the diffusion process, when jumps of lengths $\pm h, \pm 2h$ occur with equal probability (one jump per time step) and a hypothetical linear profile $c(x/h)$ is applied with fixed $c(x=0)=100$. The left part shows the jumps at the left boundary, where the postulated linear concentration profile would lead to a non-stationary situation. The middle part illustrates the flux at a location $x=12h$ (far from boundary influences), where a constant concentration gradient leads to a stationary situation. 
}
\label{f.Masterequation}
\end{figure}

Finally, we show an easy example to understand the occurence of a non-linear $c(x)$ close to the boundaries. We note that a stationary current requires $c(x,t)=\mbox{const}$ in time. Therefore at each lattice site, the number of in- and outgoing particles has to be statistically equal and constant. For illustration, we have plotted a hypothetical linear concentration $c(x)=c_0-x/L$ in Fig.~\ref{f.Masterequation} for a linear system with discrete jumps of lengths $l=\pm h$ and $\pm 2h$, each with a probability $1/4$. In accordance with the method of \cite{evans}, $c(0)$ is kept constant. For simplicity, we assume one jump per time step, as in Fig.~\ref{f.Gaussprofile}(c). 

Only for $x\geq2h$, this concentration profile leads to a constant current $j$ at all positions $x$, as a prerequisite for stationarity.
For the boundary sites, as e.g. for the site at $x=h$, however this is not true: as a consequence of the different spectrum of jump probabilities, a strictly linear $c(x)$ would not lead to a stationary current at the boundaries. Hence, close to the channel entrance, stationarity implies concentration profiles deviating from linearity as observed in Fig.~\ref{f.Gaussprofile}.  
Note that the condition of all particles starting at $x=0$ is different from the given experimental situation where particles could enter from outside the system, but it is a necessary ingredient of the method of \cite{evans}. Even when a homogeneous concentration outside the pore is applied a similar behaviour would emerge.

\section{Conclusions}

We have considered molecular diffusion in 3d channel pores with different roughness under the so-called Knudsen conditions, i.e. for negligible mutual molecular collisions and for flight times notably exceeding the periods of interaction with the pore walls and 1d systems with many different jump lengths and jump time distributions. 
In the case of the three-dimensional pores, which reflect a situation possibly occurring in the real nano-world, complete compatibility with the laws of normal diffusion is predicted. For non-interacting particles, as implied in the considered case of Knudsen diffusion, this has to lead to equivalence between transport diffusion and self-diffusion, as illustrated in \cite{kaerger92}, on the basis of Fick's and Einstein's diffusion equations. 
Assuming a constant concentration gradient, errors between $\sim5\%$ (for the three-dimensional pores) and $\sim50\%$ (for certain random walks) occur in the calculation of $D_t$ when applying the method of \cite{evans} without calculating $c(x)$ explicitely.

This complication is related to the fact that, as a typical feature of this type of simulations, for molecules entering the system only the cross section at $x=0$ is considered, while jumps out of the system may get to positions far beyond this plane. We have shown that the transmission probability $f_t$ from \cite{evans} may as well be applied to calculate the real concentration profiles, so that one is released of the computational expenses needed for the establishment of stationary conditions. With these thus calculated concentration profiles, complete 
equivalence of both diffusion coefficients $D_{s}$ and $D_{t}$ is found in all considered cases. 
Accordingly transport diffusion and self- (or tracer-) diffusion are equal with respect to both their absolute values and their dependence on the surface roughness. 
Having clarified the diffusion behaviour in single pores we have established a sound theoretical basis for the exploration of mass transfer in the numerous mesopores of nanoporous materials. This is important since in many technological applications \cite{kaerger92,chen,ertl,schueth}, it is this process of mass transfer which decides about the performance of these materials.

\acknowledgments
We gratefully acknowledge financial support by the "Deutsche Forschungsgemeinschaft" and by the "Fonds der Chemischen Industrie".


\begin{thebibliography}{0}
\bibitem{footnote1} For the results shown here we used a threshold of $0.1\%$ for the maximum concentration change between time $t$ (end of the simulation) and $t/2$. We also tested lower percentages and different time intervals, as e.g. changes between two successive time steps, but the results remained unchanged.
\end{thebibliography}
\end{document}